# Self-Repolarization process in dual-Omnipolarizers

N. Berti,[1] M. Guasoni,[2] and J. Fatome[1,*]

[1] *Laboratoire Interdisciplinaire Carnot de Bourgogne, UMR 6303 CNRS-Université de Bourgogne, Dijon, France*
[2] *Optoelectronics Research Centre, University of Southampton, Southampton SO17 1BJ, United Kingdom*
*Corresponding author: Julien.Fatome@u-bourgogne.fr*

**Abstract:** We report on an extension of the concept of nonlinear self-repolarization process by means of two different architectures based on *dual-Omnipolarizers*. More specifically, we compare the performance in terms of polarization attraction capabilities provided by two novel arrangements: The first configuration relies on two cascaded *Omnipolarizers*, whilst the second architecture integrates an additional device directly into the feedback loop. Our study reveals that for a constant power budget, the cascading of two subsequent *Omnipolarizers* enables to improve the efficiency of the attraction process, yielding an output Degree-of-Polarization close to unity, but at the cost of twofold equipments.

## 1. Introduction

Among the three independent parameters of light propagating in a monomode optical fibre, i.e. the wavelength, power and state-of-polarization (SOP), the latter remains the hardest variable to predict and master. Indeed, despite tremendous advances in many fields of photonics, especially in fibre-based systems, the light SOP remains an elusive parameter, which is still challenging to control deterministically. However, it is noteworthy that impressive advances in the manufacturing process of optical fibers have been achieved in the last decades. In particular, by means of sophisticated spinning techniques, fibre manufactures can now deliver standard telecom fibres with very weak levels of polarization mode dispersion (PMD) [1-4]. Nonetheless, the weak amount of residual birefringence combined with surrounding stress, make the polarization of light genuinely unpredictable after only a few hundreds of meters of propagation [5-8].

To overcome the impairments induced by the randomness of polarization in fibre systems, different strategies have been implemented, such as the use of polarization diversity combined with coherent detection and digital signal processing [9-10] or opto-electronic polarization tracking devices [11-13]. Nevertheless, these *scenarii* are essentially based on opto-electronic technologies, which could hinder the development of fully transparent optical networks. In this context and beyond its fundamental interest, the all-optical control of light SOP by means of nonlinear effects represents a complementary and alternative approach.

For that purpose, several attempts have been reported in the literature to realize an "ideal polarizer" enable to repolarize 100% of an arbitrary polarized lightwave — "Ideal" meaning: without polarization depending loss in stark contrast with usual Glan polarizers. In 2000, Heebner and co-workers have first proposed a "universal polarizer" capable of repolarizing an unpolarized light with more than 50% of efficiency in photorefractive materials [14]. Subsequently, the phenomenon of nonlinear polarization pulling in optical fibers has been the subject of numerous studies, relying on different types of nonlinear process such as the Raman effect [15-17], stimulated Brillouin scattering [18-19], parametric amplification [20-21] or counter-propagating four-wave mixing interactions [22-28]. According to these works, the injection of a reference pump wave is a prerequisite for the existence of a repolarization phenomenon, which thus plays the role of natural attractor for the incident lightwave. Inevitably, the drawback of these systems is that one need an ultra-stable polarized pump wave to ensure an efficient pulling effect.

In contrast with this common belief, a pump-free self-repolarization process was originally proposed in 2012 in a device called *Omnipolarizer* [29-31]. The basic principle was to introduce a feedback loop setup in the form of a reflecting element at the output of the fibre so as to force the signal to nonlinearly interact with its own backward replica through a fully-degenerate four-wave mixing interaction [29]. The origin of this self-organization effect can be understood as a direct consequence of the existence of spatial polarization domain wall solitons described in the 80s' by Zakharov and Mikhailov [32] and observed latter by Pitois *et al.* [33]. This localized spatial polarization interface is formed at the middle point of an isotropic optical fibre. It originates from the fact that for a system of two counter-propagating waves injected at both ends of an isotropic nonlinear Kerr medium, the only stable polarization configuration relies on two co-rotating circular SOPs, as first suggested in ref. [34] by Kaplan and confirmed latter in [35-36]. Any other couples of SOPs are unstable due to the non-reciprocal birefringence induced by the cross-polarization coupling occurring between the forward and backward waves. In a similar fashion, in the case of the self-repolarization process taking place within the *Omnipolarizer*, the system of two counter-propagating replicas will naturally relax towards a reciprocal stationary regime, offering a stable and fixed polarization point localized at the fibre end — at the mirror interface [29].

Based on our previous studies, here we investigate the performance of the self-repolarization phenomenon in terms of efficiency (output degree-of-polarization, DOP) and power budget for two different kinds of architectures based on *dual-Omnipolarizers*. The first configuration relies on two cascaded *Omnipolarizers*, while the second one is designed as an imbricated *Omnipolarizer* — the second device being implemented directly into the feedback loop of the main system. Our study reveals that for a constant power budget, the cascading of two subsequent *Omnipolarizers* enables to improve the efficiency of the attraction process, though requiring twice the amount of equipment's.

## 2. Principle-of-operation

The principle-of-operation of the *dual-Omnipolarizers* is illustrated in Fig. 1. The first configuration is made of a series of two cascaded independent *Omnipolarizers*. Each device consists of a segment of km-long normal dispersion optical fibre [see Table 1 for more details on the fibre parameters] encapsulated between an input optical circulator (to inject the arbitrary polarized signal and reject its backward replica) and a partially reflective element implemented at the opposite end (to send back the backward replica and collect the output signal). In this work, the feedback element consists of an EDFA-based loop, for which the amplifier gain enables to adjust the reflective coefficient beyond unity so as to operate in the monostable regime of the device [30-31]. The second configuration under study is a slight variation of the cascaded one and relies on the integration of the second *Omnipolarizer* directly into the reflecting loop of the first device. In that configuration, the feedback signal now represents a "cleaner" replica of the incident lightwave. However, it is important to note that the insertion of km-long fibre segments into the reflecting loop could potentially induce more complex and chaotic behaviours [37]. Indeed, the large delay imposed by the time of flight into the reflecting loop can introduce a decorrelation between typical SOP fluctuations and the response-time of the *Omnipolarizer* — essentially set by the usual nonlinear length $L_{nl} = 1/\gamma P$, where $\gamma$ corresponds to the nonlinear Kerr coefficient and $P$ the injected power [38].

|  | L km | α dB/km | γ W$^{-1}$.km$^{-1}$ | PMD ps/km$^{1/2}$ |
|---|---|---|---|---|
| Fibre #1 | 6.2 | 0.2 | 1.7 | < 0.05 |
| Fibre #2 | 10 | 0.2 | 2.1 | < 0.05 |

**Table 1.** Fibre parameters involved in the *dual-Omnipolarizers*. L: fibre length, α: fibre losses, γ: Kerr coefficient, PMD: Polarization mode dispersion.

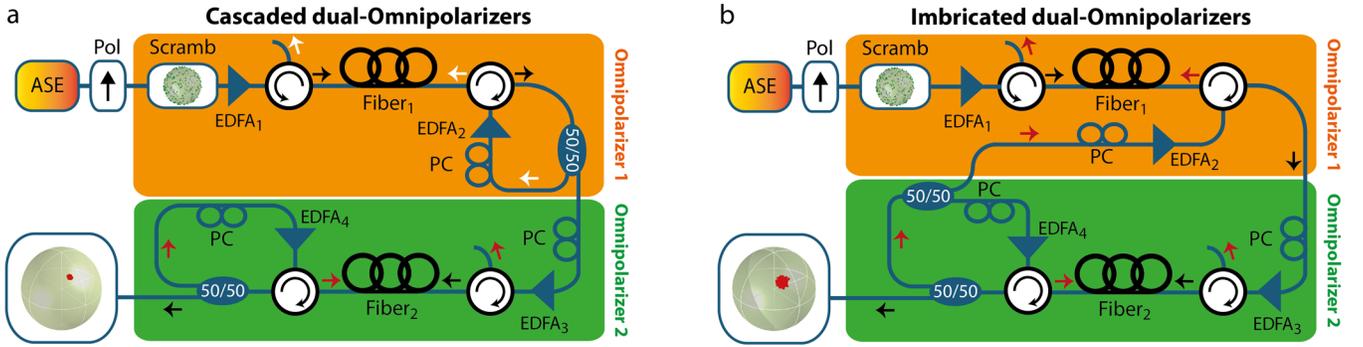

**Fig. 1.** Principle-of-operation of the *dual-Omnipolarizers*. (a) Experimental setup consisting of two cascaded *Omnipolarizers*. (b) Experimental setup involving two imbricated *Omnipolarizers*. PC: Polarization controller, EDFA: Erbium doped fibre amplifier, CIR: Optical circulator, Pol: Polarizer, ASE: Amplified spontaneous emission, Scramb: Opto-electronic polarization scrambler.

## 3. Modelling

In order to predict the spatio-temporal dynamics of both counter-propagating waves in the different configurations of *dual-Omnipolarizer* understudy, it appears convenient to introduce the Stokes formalism. To this aim, we indicate with $S = [S_1, S_2, S_3]$ and $J = [J_1, J_2, J_3]$ the Stokes vectors associated with the forward and backward waves of power $S_0$ and $J_0$, respectively. Note that, here the north and south poles of the Poincaré sphere correspond respectively to the right and left circular SOPs, i.e. $S_3 = J_3 = \pm 1$. Using these notations, in a randomly birefringent fibre, like those used in the present experiments, the evolution of the SOPs of the forward and backward waves in each *Omnipolarizer* can be described by the following set of two coupled partial differential equations [39]:

$$\begin{cases} v^{-1}\partial_t S + \partial_z S = S \wedge DJ - \alpha S \\ v^{-1}\partial_t J - \partial_z J = J \wedge DS - \alpha J \end{cases}, \quad (1)$$

where $D = 8/9\gamma\, diag(-1, -1, +1)$ is a diagonal matrix [39], $\wedge$ denotes the vector product, $z$ the propagation distance, $\alpha$ the fibre losses and $v$ corresponds to the group velocity of the waves in the fibre. Note that we are only interested in the spatio-temporal dynamics of continuous-wave SOPs so chromatic dispersion has been neglected. Despite its simplicity, Eqs. (1) are able to reproduce all the essential features of the polarization dynamics observed in the *Omnipolarizers*, which is highlighted in the results section by the good agreement obtained between numerical and experimental results. Additionally, this set of equations has to be completed by the boundary conditions imposed by the reflecting element such that at any time $t$, for individual devices and the cascaded configuration: $J^{(i)}(z = L^{(i)}, t) = \rho S^{(i)}(z = L^{(i)}, t)$, where the superscript $(i)$ denotes the Omnipolarizer ($i$ ={1,2}), $L^{(1)} = 6.2$ km and $L^{(2)} = 10$ km indicate respectively the fiber length in the Omnipolariser 1 and 2; $\rho$ denotes the gain coefficient of the reflecting loop in the *Omnipolarizers* — hereafter $\rho = 2$ in the experimental section. Moreover, for the cascaded architecture, the boundary conditions also impose: $S^{(2)}(z = 0, t) = S^{(1)}(z = L^{(1)}, t)$. In contrast, for the imbricated version, these conditions read: $J^{(1)}(z = L^{(1)}, t) = J^{(2)}(z = L^{(2)}, t) = \rho S^{(2)}(z = L^{(2)}, t)$.

## 4. Results

### 1. Characterization of individual Omnipolarizers

Firs-of-all, we characterized the performance of both individual *Omnipolarizers* based on fibres #1 and #2 indicated in Table 1. To this aim, we have injected an arbitrary polarized signal and evaluated its output degree-of-polarization (DOP) with respect to the injected power as followed [38-39]:

$$\text{DOP} = \frac{\sqrt{\langle S_1(t)\rangle^2 + \langle S_2(t)\rangle^2 + \langle S_3(t)\rangle^2}}{\langle S_0\rangle} \quad (2)$$

For these measurements, the incoming signal consists of an arbitrary polarized, 100-GHz bandwidth, partially-coherent wave centred around 1550 nm. This signal is generated by means of a spectrally-sliced, Erbium-based, amplified spontaneous-noise-emission source (ASE) followed by a polarizer (see Fig. 1). This large bandwidth incident signal has the benefit to avoid any growth of stimulated Brillouin scattering in the fibre under-test, which could occur due to the large levels of power involved in our experiments (typically beyond 0.5 W). The resulting signal is then randomly polarized thanks to a commercial scrambler at a speed of 0.5 kHz and a sequence made of 256 different SOPs — see the input Poincaré sphere reported in Fig. 1a and 3a. Figure 2 summarizes these measurements. Panel (a) depicts the DOP of the forward signal evaluated at the output of the *Omnipolarizer* #1 (blue) and #2 (red) as a function of the total input power — defined as $S_0 + J_0$ — and for a reflecting gain coefficient $\rho = 2$. We can first emphasize that, as expected, both equivalent devices exhibit similar dynamics. Then we can observe that both *Omnipolarizers* are able to repolarize the initial scrambled signal with an efficiency close to 100% for a total injected power beyond 1 W. To complete this analysis, we have reported in Fig. 2b the average interspace evaluated in-between each couple of points of the output Poincaré sphere (exemplified in Fig. 3b) with respect to the input power. These results show that both the relative inter-distance and its associated variance are strongly reduced by the self-attraction process, thus minimizing the basin of attraction onto the Poincaré sphere. Once again, since based on the same architecture, both devices exhibit very similar behaviours.

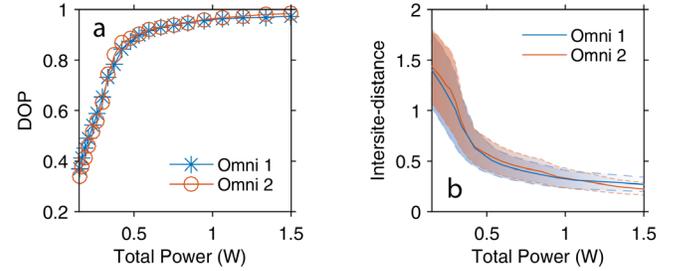

**Fig. 2.** (a) Degree-of-Polarization (DOP) of the repolarized wave evaluated at the output of the *Omnipolarizer* #1 (blue) and #2 (red) as a function of the total injected power ($S_0 + J_0$). (b) Average intersite-distance (solid lines) between two points of the Poincaré sphere measured at the output of the *Omnipolarizer* #1 (blue) and #2 (red) with respect to the total injected power. The respective shaded areas display the corresponding variance.

### 2. Characterization of the dual-Omnipolarizers

The performance of a single *Omnipolarizer* was then compared to both configurations described in Fig. 1 and involving different arrangements of two devices. To this aim, we fixed a constant power budget of 1.5 W — sum of all injected power — and recorded the resulting Poincaré sphere at the output of each configuration. Figure 3 summarizes our observations. At first, Fig. 3a shows the polarization of the incoming signal and reveals that the initial scrambling process efficiently randomizes the input SOP, clearly illustrated by the uniform coverage of the Poincaré sphere. In stark contrast, Fig. 3b displays the resulting SOP fluctuations recorded at the output of the single *Omnipolarizer* #1 for a total power of 1.5 W. As inferred by Figs. 2a&b, all the output SOPs have now converged around a small area of the Poincaré sphere, confirming that a randomly polarized lightwave can spontaneously self-organize its state-of-polarization within the *Omnipolarizer* [29]. Figures 3(c&d) report similar measurements for both arrangements of *dual-Omnipolarizers*. Figure 3c demonstrates that the efficiency of the repolarization process can be significantly improved by the transmission of the incident signal through two successive devices, resulting in a very small pool of attraction onto the output Poincaré sphere. Similarly, the imbricated configuration, reported in Fig. 3d and for which the feedback replica represents a "cleaner" version of the forward signal, also allows to notably reduce the residual fluctuations observed in output of the single device. But at this stage, it is still difficult to settle which of the dual-configurations provides the better performance in terms of repolarization.

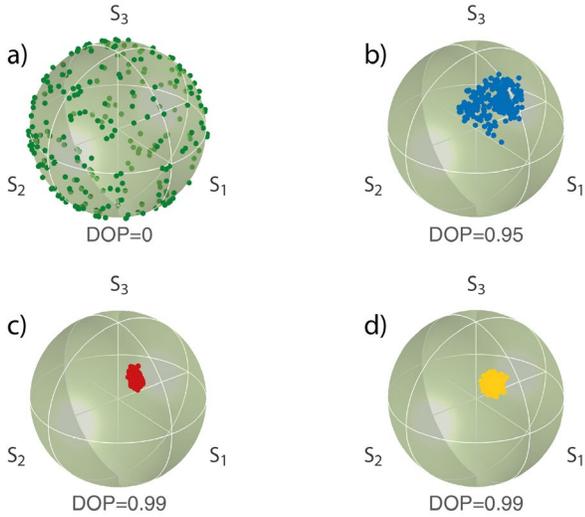

**Fig. 3.** (a) Poincaré unit sphere recorded at the input of the system. (b) Resulting Poincaré sphere measured at the output of a single *Omnipolarizer* #1. (c) Poincaré sphere measured after two cascaded *Omnipolarizers* 1&2. (d) Poincaré sphere measured at the output of the imbricated *dual-Omnipolarizer*. Note that all the spheres have been recorded for a constant power budget of 1.5 W.

To further assess the performance of the three configurations under study, we have reported in Fig. 4a the DOP measured at the output of the three arrangements as a function of the total injected power. From a general point of view, all the three configurations lead to an efficient repolarization process of the incident scrambled signal. We can however highlight the superiority of the cascaded version of two *Omnipolarizers*, which allows a faster repolarization dynamic for a constant power budget, while surpassing the capacity of a single device (limited to 1.5 W) for larger levels of power, reaching an output DOP very close to unity. Regarding the imbricated *dual-Omnipolarizer*, it does not allow to improve the performance of the repolarization process for moderate average power, but enables to reach higher values of DOP than the single configuration for larger budgets of power, though essentially due to the use of more EDFAs. Furthermore, this configuration has been found more sensitive to the adjustment of the different polarization controllers than the cascaded version. To go deeper into the analysis, Fig. 4b displays the average intersite-distance evaluated on the output Poincaré sphere for the 3 configurations described in Panel (a) as a function of the injected power. These results clearly show that the cascaded version of *dual-Omnipolarizers* provides the best performance with a higher convergence of the SOPs onto the Poincaré sphere for a constant power budget, whilst surpassing the single-device configuration above its maximum power capability. This conclusion is not surprising since the additional *Omnipolarizer* always acts as a second stage of repolarization and thus benefits from a highly favourable and pre-processed input signal. Finally, Figs. 4(c-d) display the corresponding numerical simulations based on Eqs. (1) and show a reasonable qualitative agreement with our experimental results reported in Panels (a&b), in particular confirming the efficiency of the systems based on two *Omnipolarizers*.

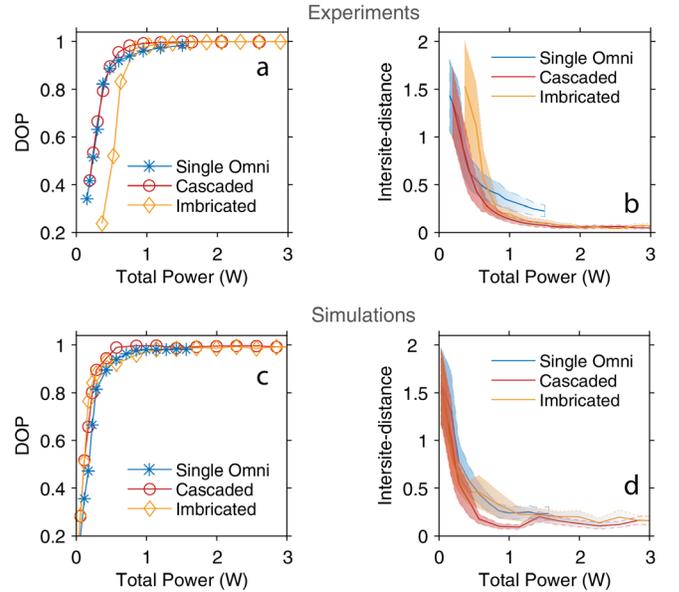

**Fig. 4.** (a) Degree-of-Polarization (DOP) with respect to the total input power recorded at the output of the three different configurations of *Omnipolarizers*: single device (blue), two cascaded *Omnipolarizers* (red) and imbricated *Omnipolarizers* (yellow). (b) Average intersite-distance (solid lines) and standard deviation (shaded areas) between two points of the Poincaré sphere as a function of power for the three configurations understudy. (c) & (d) Corresponding numerical simulations based on Eqs. (1).

## 5. Conclusion

To summarize, in this work we have reported on an extension of the concept of nonlinear self-repolarization phenomenon by means of several configurations based on *dual-Omnipolarizers*. More precisely, we have compared the improvement in terms of polarization attraction capabilities provided by two novel arrangements: first, the cascading of two *Omnipolarizers* and secondly, an original version of imbricated *dual-Omnipolarizers* for which the feedback replica self-organizes its SOP in an additional nested device. We have concluded that, for a constant power budget, the cascading of two subsequent devices provides the best performance and enables a more efficient repolarization process than a single device, whilst reducing further the output fluctuations for larger injected powers, and reaching DOPs close to unity. Nevertheless, these improvements have been obtained at the cost of a more complex and expensive implementation. To conclude, the generalization of this work and the cascading of *N-Omnipolarizers* could overcome the limits imposed on the fibre length and on the injected power but on the other hand would dramatically increase the complexity of the cascading setup — proportional to the number of devices — which practically puts a limit on the maximum number of subsequent stages.

**Funding.** J.F. acknowledges financial support from the CNRS, the Conseil Régional de Bourgogne Franche-Comté and the European Union through the PO FEDER-FSE Bourgogne 2014/2020 programs.

**Acknowledgment**. The authors thank Dr A. Picozzi for fruitful discussions.

## REFERENCES

1. T. Geisler, "Low PMD Transmission Fibres," in European Conference on Optical Communications ECOC 2006, paper Mo.3.3.1, 2006.
2. A. J. Barlow, J. J. Ramskov-Hansen, and D. N. Payne "Birefringence and polarization mode-dispersion in spun singlemode fibers," Appl. Opt., vol. 20, pp. 2962-2968, 1981.
3. M. J. Li and D. A. Nolan, "Fiber spin-profile designs for producing fibers with low polarization mode dispersion," Opt. Lett., vol. 23, pp. 1659-1661, 1998.
4. L. Palmieri, "Polarization Properties of Spun Single-Mode Fibers," IEEE J. Lightw. Technol., vol. 24, pp. 4075-4088, 2006.
5. M. Boroditsky, M. Brodsky, N. J. Frigo, P. Magill, and H. Rosenfeldt, "Polarization dynamics in installed fiber optic systems" IEEE LEOS Annual Meeting Conference Proceedings (LEOS), pp. 413–414, 2005.
6. P. M. Krummrich and K. Kotten, "Extremely fast (microsecond scale) polarization changes in high speed long hail WDM transmission systems," in Proc. of Optical Fiber Commun. Conference, Los Angeles, USA, 2004.
7. S. C. Rashleigh, "Origins and control of polarization effect in single-mode fibers," IEEE J. Lightw. Technol., vol. 1, pp. 312–331, 1983.
8. A. Simon and R. Ulrich, "Evolution of polarization along a single-mode fiber," Appl. Phys. Lett., vol. 31, pp. 517–520, 1977.
9. J. Renaudier, G. Charlet, M. Salsi, O. B. Pardo, H. Mardoyan, P. Tran, and S. Bigo, "Linear Fiber Impairments Mitigation of 40-Gbit/s Polarization-Multiplexed QPSK by Digital Processing in a Coherent Receiver," IEEE J. Lightwave Technol., vol. 26, pp. 36-42, 2008.
10. G. Charlet, J. Renaudier, M. Salsi, H. Mardoyan, P. Tran, and S. Bigo, "Efficient Mitigation of Fiber Impairments in an Ultra-Long Haul Transmission of 40Gbit/s Polarization-Multiplexed Data, by Digital Processing in a Coherent Receiver," in Optical Fiber Communication Conference and Exposition, paper PDP17, 2007.
11. M. Martinelli, P. Martelli, and S. M. Pietralunga, "Polarization stabilization in optical communications systems," IEEE J. Lightw. Technol., vol. 24, pp. 4172–4183, 2006.
12. B. Koch, R. Noe, D. Sandel, and V. Vitali Mirvoda, "Versatile endless optical polarization controller/tracker/demultiplexer," Opt. Express vol. 22, pp. 8259–8276, 2014.
13. W. H. J. Aarts and G. Khoe, "New endless polarization control method using three fiber squeezers," IEEE J. Lightwave Technol., vol. 7, pp. 1033-1043, 1989.
14. E. Heebner, R. S. Bennink, R. W. Boyd, and R. A. Fisher "Conversion of unpolarized light to polarized light with greater than 50% efficiency by photorefractive two-beam coupling," Opt. Lett., vol. 25, pp. 257-259, 2000.
15. M. Martinelli, M. Cirigliano, M. Ferrario, L. Marazzi, and P. Martelli, "Evidence of Raman-induced polarization pulling," Opt. Express, vol. 17, pp. 947-955, 2009.
16. L. Ursini, M. Santagiustina, and L. Palmieri, "Raman Nonlinear Polarization Pulling in the Pump Depleted Regime in Randomly Birefringent Fibers," IEEE Photon. Technol. Lett., vol. 23, pp. 1041-1135, 2011.
17. N. J. Muga, M. F. S. Ferreira, and A. N. Pinto "Broadband polarization pulling using Raman amplification," Opt. Express, vol. 19, pp. 18707-18712, 2011.
18. L. Thevenaz, A. Zadok, A. Eyal and M. Tur, "All-optical polarization control through Brillouin amplification", in Optical Fiber Communication Conference, OFC'08, paper OML7, 2008.
19. Z. Shmilovitch, N. Primerov, A. Zadok, A. Eyal, S. Chin, L. Thevenaz, and M. Tur "Dual-pump push-pull polarization control using stimulated Brillouin scattering," Opt. Express, vol. 19, pp. 25873-25880, 2011.
20. B. Stiller, P. Morin, D. M. Nguyen, J. Fatome, S. Pitois, E. Lantz, H. Maillotte, C. R. Menyuk, and T. Sylvestre, "Demonstration of polarization pulling using a fiber-optic parametric amplifier," Opt. Express, vol. 20, pp. 27248-27253, 2012.
21. M. Guasoni, V. Kozlov, and S. Wabnitz, "Theory of polarization attraction in parametric amplifiers based on telecommunication fibers," J. Opt. Soc. Am. B, vol. 29, pp. 2710-2720, 2012.
22. S. Pitois and M. Haelterman, "Optical fiber polarization funnel," Nonlinear Guided Waves and Their Applications NLGW'01, paper MC79-1, 2001.
23. J. Fatome, S. Pitois, P. Morin, and G. Millot, "Observation of light-by-light polarization control and stabilization in optical fibre for telecommunication applications," Opt. Express, vol. 18, pp. 15311-15317, 2010.
24. K. Turitsyn and S. Wabnitz, "Stability analysis of polarization attraction in optical fibers," Opt. Commun., vol. 307, pp. 62-66, 2013.
25. V. V. Kozlov, J. Fatome, P. Morin, S. Pitois, G. Millot and S. Wabnitz, "Nonlinear repolarization dynamics in optical fibers: transient polarization attraction," J. Opt. Soc. Am. B, vol. 28, pp. 1782-1791, 2011.
26. M. Barozzi and A. Vannucci "Lossless polarization attraction of telecom signals: application to all-optical OSNR enhancement," J. Opt. Soc. Am. B, vol. 31, pp. 2712-2720, 2014.
27. V. Costa Ribeiro, R. S. Luis, J. M. D. Mendinueta, B. J. Puttnam, A. Shahpari, N. J. C. Muga, M. Lima, S. Shinada, N. Wada and A. Teixeira "All-Optical Packet Alignment Using Polarization Attraction Effect," IEEE Photon. Technol. Lett., vol. 27, pp. 541-544, 2015.
28. A. DeLong, W. Astar, T. Mahmood and G. M. Carter, "Polarization attraction of 10-Gb/s NRZ-BPSK signal in a highly nonlinear fiber," Opt. Express vol. 25, pp. 25625-25636, 2017.
29. J. Fatome, S. Pitois, P. Morin, D. Sugny, E. Assémat, A. Picozzi, H. R. Jauslin, G. Millot, V. V. Kozlov and S. Wabnitz, "A universal optical all-fiber omnipolarizer," Sci. Rep., vol. 2, 938, 2012.
30. P. Y. Bony, M. Guasoni, P. Morin, D. Sugny, A. Picozzi, H. Jauslin, S. Pitois and J. Fatome, "Temporal spying and concealing process in fibre-optic data transmission systems through polarization bypass," Nat. Commun., vol. 5, pp. 5:4678, 2014.
31. P.-Y. Bony, M. Guasoni, E. Assémat, S. Pitois, D. Sugny, A. Picozzi, H. R. Jauslin and J. Fatome, "Optical flip-flop memory and data packet switching operation based on polarization bistability in a telecomunnication optical fiber," J. Opt. Soc. Am. B, vol. 30, pp. 2318-2325, 2013.
32. V. E. Zakharov and A. V. Mikhailov, "Polarization domains in nonlinear optics," JETP Lett., vol. 45, pp. 349-352, 1987.
33. S. Pitois, G. Millot, and S. Wabnitz, "Polarization domain wall solitons with counterpropagating laser beams," Phys. Rev. Lett., vol. 81, pp. 1409-1412, 1998.
34. A. E. Kaplan, "Light-induced nonreciprocity, field invariants, and nonlinear eigenpolarizations," Opt. Lett., vol. 8, pp. 560-562, 1983.
35. S. Wabnitz and G. Gregori, "Symmetry-breaking and intrinsic polarization instability in degenerate four-wave mixing," Opt. Commun. Vol. 59, pp. 72-76, 1986.
36. A. L. Gaeta, R. W. Boyd, J. R. Ackerhalt and P. W. Milonni, "Instabilities and chaos in the polarization of counterpropagating light fields," Phys. Rev. Lett., vol. 58, pp. 2432-2435, 1987.
37. J. Morosi, N. Berti, A. Akrout, A. Picozzi, M. Guasoni and J. Fatome, "Polarization chaos and random bit generation in nonlinear fiber optics induced by a time-delayed counter-propagating feedback loop," Opt. Express, vol. 26, pp. 845-858, 2018.
38. G. P. Agrawal, Nonlinear Fiber Optics, 4th ed., Academic Press, New York, 2007.
39. V. V. Kozlov, J. Nuno and S. Wabnitz, "Theory of lossless polarization attraction in telecommunication fibers," J. Opt. Soc. Am. B, vol. 28, pp. 100-108, 2011.